\documentclass[aps,prb,twocolumn,fleqn]{revtex4}
\usepackage[latin1]{inputenc}
\usepackage{amsmath}
\usepackage{amsfonts}
\usepackage{amssymb}

\usepackage{graphicx} 
\def\gapx{\lower 2pt \hbox{$\buildrel>\over{\scriptstyle{\sim}}$\ }}
\def\lapx{\lower 2pt \hbox{$\buildrel<\over{\scriptstyle{\sim}}$\ }}
\def\ph2{{\it p}-H$_2$}
\def\beq{\begin{equation}}
\def\eeq{\end{equation}}
\def\Am3{\AA$^{-3}$}
\begin{document}

\widetext
\title{Quasi-1D parahydrogen in nanopores}
\author{Tokunbo Omiyinka and Massimo Boninsegni   } 
\affiliation{Department of Physics, University of Alberta, Edmonton, Alberta, Canada T6G 2G7}
\date{\today}

\begin{abstract}
The low temperature physics of parahydrogen (\ph2) confined in cylindrical channels of diameter of the order of 1 nm  is studied theoretically by Quantum Monte Carlo simulations.  On varying the attractive strength of the wall of the cylindrical pore, as well as its diameter,  the equilibrium phase evolves from  a single quasi-1D channel 
along the axis, to  a concentric cylindrical shell.  It is found that  the quasi-1D system retains a strong propensity to crystallization, even though on weakly attractive substrates quantum fluctuations reduce somewhat such a tendency compared to the purely 1D system. No evidence of a topologically protected superfluid phase (in the Luttinger sense) is observed. Implications on the possible existence of a bulk superfluid phase of parahydrogen are discussed.

\end{abstract}
\maketitle
\section{INTRODUCTION}
The speculative superfluid phase of condensed parahydrogen (\ph2), predicted over forty years ago,\cite{ginzburg} has so far eluded a direct, unambiguous experimental observation. The problem with the original prediction is  that it assumes fluidlike behaviour of \ph2 at low $T$, when in fact the system undergoes crystallization (at saturated vapor pressure) at $T=13.8$ K, over a factor two higher than the temperature ($\sim$ 6 K) at which Bose-Einstein Condensation and superfluidity ought to occur.\cite{clark} Solidification takes place  due to the depth of the attractive well of the intermolecular potential, roughly three times that between two helium atoms. The zero temperature equilibrium phase of \ph2 is theoretically predicted to be a (non-superfluid) crystal  in reduced dimensions as well.\cite{myprof3,myprof4} 
\\ \indent
First principle theoretical calculations based on realistic intermolecular potential\cite{sindzingre,fabio,fabio2}  yield evidence  that small \ph2 clusters (thirty molecules or less) remain liquidlike down to temperatures of the order of 1 K, where they display superfluid behavior. This lends some credibility to the hypothesis that a bulk superfluid phase might be observable, if crystallization of the fluid could be suppressed. One way of lowering the freezing temperature of a fluid consists of confining it in a porous medium such as vycor glass.\cite{tell, molz,notemink,diso} 
In the pores of vycor, whose characteristic size\cite{mason} is $\sim 4$ nm, \ph2 freezes\cite{sokol} at a temperature $T \sim$ 8 K,  considerably lower  than that of bulk \ph2, but still significantly higher than the estimated superfluid transition temperature; indeed,
the search for superfluid behaviour of \ph2 in vycor has not met with success.\cite{bretz81,schindler96} The solid phase, which has a different crystal structure than that of bulk \ph2, nucleates at the pore walls;\cite{sokol}
one may therefore wonder whether different physics (possibly including a superfluid phase) may be observed in a different confining environment, e.g., with pores 
of smaller size and/or  less strongly adsorbing than silica. 
\\ \indent
It was recently shown\cite{enhanced} that  \ph2 clusters of $\sim 30$ molecules, confined in  weakly adsorbing spherical cavities of size $\sim 2$ nm,  surprisingly feature an enhanced superfluid response, compared to that of free clusters. It is therefore conceivable that the same effect may take place in a different geometry, e.g., cylindrical channels,  experimentally relevant to actual porous media. 
As the diameter of the channel is reduced to one nm or less, the physics of a fluid confined in it approaches the 1D limit, and becomes therefore amenable to interpretation within the framework of Luttinger liquid theory (LLT),\cite{haldane} as shown by computer simulation  of $^4$He confined in nanopores of such a characteristic size.\cite{HeLutt}\\ \indent
No true long range order exists in 1D; rather, two-body 
correlation functions display slow power-law decays at long distances. Of particular interest in this work is the pair correlation function $g(r)$, which at any finite temperature $T$ will take on the following behavior in the thermodynamic limit:
\begin{equation}\label{maineq}
g(r)\sim 1-\frac {1}{2\pi^2 K\rho^2 r^2}+A\  {\rm cos}(2\pi\rho r)\frac {1} {\rho^{2} r^{2/K}}
\end{equation}
Here, $\rho$ is the linear density of particles and $A$ a non-universal, system-dependent constant. The Luttinger parameter $K$ 
describes how quickly the oscillations of $g(r)$ decay at long distance;  based on its value   one can  meaningfully differentiate between phases that are ``quasi-crystalline" or ``quasi-superfluid" in character. Specifically, 
if $K > 2$ the static structure factor will develop (Bragg) peaks at reciprocal lattice vectors, which is the experimental signature of a crystalline solid. On the other hand, if $K < 0.5$ the system features
 a robust propensity to superflow, and there exists a well-defined theoretical scenario in which a 3D bulk superfluid phase may arise in a network of interconnected, quasi-1D channels.\cite{shevchenko,dislocation,stan} For $0.5 \le K \le 1.5$, the system can be regarded 
as a ``glassy" insulator (at the high end of the interval), or as a weak superfluid, subjected to pinning by either disorder or an external potential. Henceforth, whenever speaking of ``crystal" versus ``superfluid", we shall refer to the above classification.
\\ \indent
The ground state phase diagram of \ph2 in one dimension only features a crystalline phase, with a value of $K\sim 3.5$ at the equilibrium density $\rho_0=0.218$ \AA$^{-1}$, monotonicaly increasing with density and remaining above $\ge 2$ all the way down to the spinodal density.\cite{myprof4} This low temperature phase diagram in 1D qualitatively mimics that of bulk \ph2 in two and three dimensions, with no evidence of a superfluid phase.
\\ \indent
In this work, we study theoretically the phase diagram of a quasi-1D fluid of \ph2 in the narrow confines of a cylindrical channel of diameter of the order of 1 nm. The hypothesis being tested here is whether quantum excursions of molecules in the direction perpendicular to the axis of the cylinder may act to screen the intermolecular potential, to the point of reducing the propensity of the system to form a crystal, possibly stabilizing a superfluid (in the Luttinger sense) phase.
\\ \indent
We carried out first principle computer simulations at low $T$ of a reasonably realistic model of the system of interest, essentially the same model utilized in similar studies\cite{HeLutt,glyde} of $^4$He. We utilized different sets of potential parameters, in order to impart to the wall of the channel different adsorption properties, going from a substrate as strongly attractive as silica,  to a much weaker one, such as Cs. Most of the results presented here are for channels of diameter $d$=10 \AA, but we also performed simulations with $d$=5 and 20 \AA.
\\ \indent
Inside the narrower channel, an adsorbate is thermodynamically stable (i.e., there exists a bound state) only for the most strongly adsorbing substrate considered here; in such a tight confinement, molecules remain very close to the axis of the channel, and 
the system closely reproduces the physics of purely 1D \ph2, with the same equilibrium (linear) density and the same value of the Luttinger parameter, within the uncertainty of our calculation. On the other hand, if the channel has a 2 nm diameter the equilibrium phase is consistently an insulating solidlike film adsorbed on the wall; growth takes place through successive solidlike layers, much like on a flat substrate.
\\ \indent
Considerably more diversity of behavior is observed in channels of diameter 1 nm, once again pointing to the fact that this is the characteristic length scale within which  \ph2   displays the most interesting physics. 
For the weakest substrate considered in this work (Cs), \ph2 forms inside the channel  a quasi-1D adsorbate along the axis, always on the crystalline side of the LLT but with a reduced value of $K$ (2.7) with respect to the purely 1D case. 
As the strength 
of the substrate is gradually increased, molecules experience a greater pull toward the wall,  their excursions away 
from the axis become increasingly significant, and the value of $K$ decreases, attaining a minimum (close to 2) on a Li substrate, for which
the quasi-1D fluid takes on a helical structure. On stronger substrate, the system forms at equilibrium a shell coating the walls, the 
central region of the channel remaining empty. On increasing the density, a quasi-1D second layer eventually forms, again with physical properties that mimic those of the strictly 1D system.
\\ \indent
The main physical conclusion of this investigation is that although nanoscale confinement in a  cylindrical geometry can reduce its strong tendency to crystallization, nonetheless a regime of possible superfluidity of \ph2 is never approached. This provides additional support to the notion that superfluidity in \ph2 may only be observed in finite clusters.
\\ \indent
The remainder of this paper is organized as follows: in Sec. \ref{mc} we introduce the model and provide computational details; in   Sec. \ref{sere} we illustrate our results and offer our theoretical interpretation. Finally, we outline our conclusions in Sec. \ref{conc}.
\section{model and methodology}\label{mc}
Our system of interest is modeled as an ensemble of $N$ \ph2 molecules, regarded as pointlike particles of spin zero, hence 
obeying Bose-Einstein statistics. Molecules are confined inside 
a cylindrical channel of diameter $d$ and length $L$, whose axis is in the $z$ direction, along which periodic boundary conditions are
imposed. The value of $L$ utilized in most of the calculations is 160 \AA, which we have empirically  established to be sufficient to capture quantitatively the physical behaviour inside an infinitely long channel.\\ \indent 
The many-body Hamiltonian of the system is the following:
\begin{equation}\label{eq2}
\hat H = -\lambda\sum_i \nabla_i^2 + \sum_{i<j}v(r_{ij}) + \sum_i V({\bf r}_i).
\end{equation}
Here, $\lambda=12.031$ K\AA$^{-2}$, $r_{ij}\equiv |{\bf r}_i-{\bf r}_j|$ is the distance between any two \ph2 molecules and ${\bf r}_i$ is the position of the $i$th
molecule. The potential $v$ describes the pair-wise interaction of  \ph2 molecules, whereas $V$ 
that of \ph2 molecules with the wall of the channel.  We used for $v$ a potential recently proposed by Moraldi,\cite{moraldi} modified at short distances as described in Ref. \onlinecite{pair_pot}. This potential has been shown to reproduce rather accurately the experimental equation of state of solid \ph2, up to megabar pressure.  Its use is motivated by the fact that the density of \ph2 inside the channel is not uniform, and it could be quite high in the vicinity of the surface, in case of a strongly attractive substrate; it makes therefore sense to utilize a pair potential that is more accurate at high pressure than, for example, the standard Silvera-Goldman pair potential.
\\ \indent
For the the interaction $V$ between a \ph2 molecule and the wall, we use the expression arising from the integration of a Lennard-Jones potential over an infinite continuous medium surrounding the cylinder, 
regarded as infinitely long.\cite{jcp}
\begin{figure}[h]
\centering
\includegraphics[height=0.75\columnwidth]{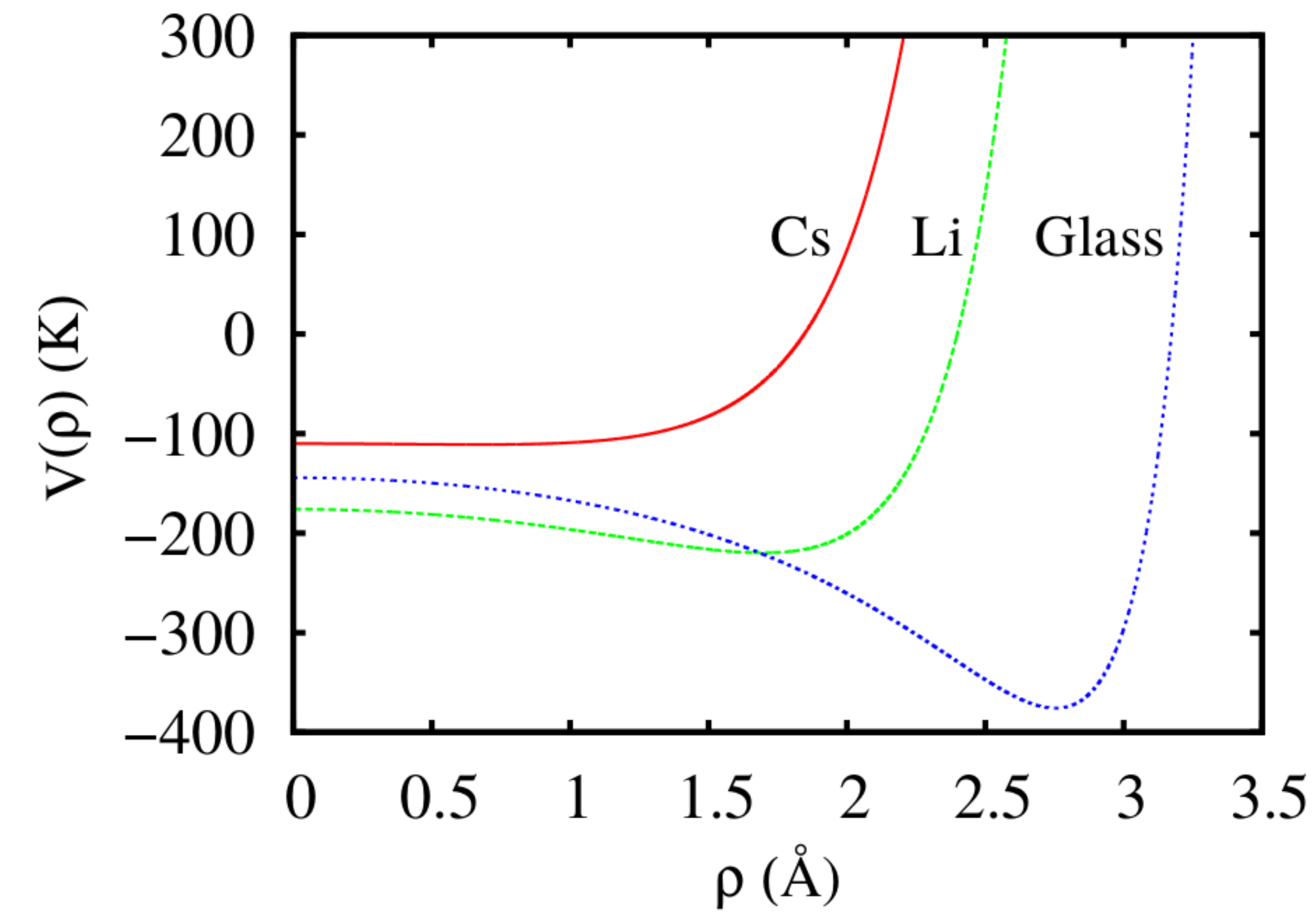}
\caption{{\it Color online}. Potentials describing the interaction between a \ph2 molecule and the wall of the confining cylindrical channel, for some of the substrates considered here. Cs (cesium) is the weakest, Glass is the most strongly attractive, and Li (lithium) has intermediate strength. Here, $\rho$ is the distance of a particle from the axis of the channel.} \label{pone}
\end{figure}
\begin{table}[h]
\begin{tabular}{|c|c|c|} \hline 
Substrate&$D$ (K)&$a$ (\AA)  \\
Cs & 37.8 & 3.88 \\
Rb & 39.7	& 3.87 \\
K & 44.4 & 3.76 \\
Na &69.5 &3.40 \\
Li & 99.7 & 3.19 \\
Mg & 191.8 & 2.76\\
Glass & 232 & 2.22 \\ \hline
\end{tabular}
\caption {\noindent {Potential parameters $D$ and $a$ utilized in this work. The values are taken from Ref. \onlinecite{chizmeshya}. Substrates are listed in order of increasing strength, from top to bottom. }
}\label{params}
\end{table}
The resulting expression only depends on the distance $\rho$ of the particle from the axis of the cylinder and contains two parameters, a characteristic length $a$ which is  essentially  the distance of closest approach to the wall of the channel, where the potential is strongly repulsive, and $D$, which has the dimensions of an energy and is basically the depth of the attractive well of the potential which particles experience from the wall.\cite{expla} These two parameters are adjusted to reproduce, as accurately as allowed by such a relatively simple model, the interaction of a particle with a substrate. The values of the parameters of the potential utilized in this work are reported in Table \ref{params}. They were taken from Ref. \onlinecite{chizmeshya}, except those for Glass, which were obtained starting from those proposed in Ref. \onlinecite{treiner} for Helium atoms near a glass substrate, using Lorentz-Berthelot mixing rule to adjust them for \ph2. Our focus here is mosty on metallic alkali substrates which are known to be weak adsorbers, for the purpose of identifying physical conditions in which the equilibrium phase of the system is quasi-1D, i.e., with \ph2 molecules lined up along the axis of the channel.
\\ \indent
We investigated the low temperature thermodynamics of our system of interest by means of first principle computer simulations based
on the worm algorithm in the continuous-space path integral representation.\cite{worm1,worm2} This  methodology allows one to calculate finite temperature thermodynamic properties of Bose systems, with the Hamiltonian as the only required input and without any uncontrolled approximations. In particular, it grants one access to energetics as well as superfluid properties (both global and local\cite{io,adrian2}) and relevant correlation functions, such as the reduced pair correlation function for the central quasi-1D channel along the axis of the cylindrical channel. As mentioned in the Introduction, we performed most of our simulations for a system with diameter $d$=10 \AA, but we also carried out a few for  a narrower channel, with diameter $d$=5 \AA, as well as a wider one, with diameter $d$=20 \AA.
\\ \indent
The calculation performed here is standard, and we refer the readers interested in the methodology to the original references. We simply mention that all of the calculations were performed with the usual short-time propagator accurate to fourth order in the imaginary time step $\tau$, and with a  value of $\tau=1/640$ K$^{-1}$, which in a few cogent test cases was found to yield estimates indistinguishable from those extrapolated to the $\tau\to 0$ limit.
\section{RESULTS}\label{sere}
\begin{figure}[h]
\centering
\includegraphics[height=0.65\columnwidth]{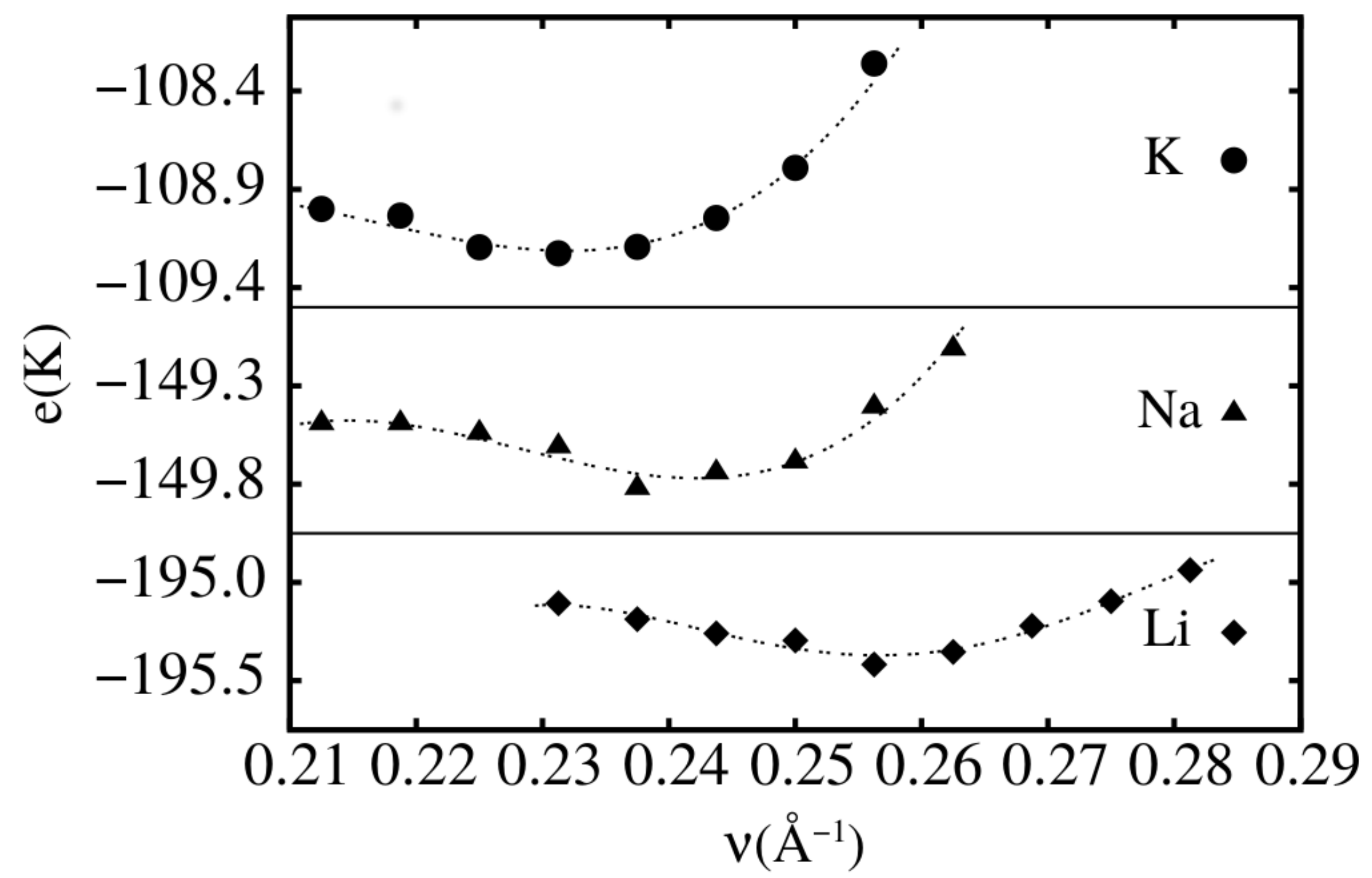}
\caption{Energy per \ph2 molecule $e$ (in K) versus linear density $\nu$ (in {\AA}$^{-1}$), at $T = 1$ K, inside a cylindrical channel of diameter $d$=10 \AA. Results are shown for three different substrates, namely K (filled circles), Na (filled triangles) and Li (diamonds), listed in order of increasing strength (see Table \ref{params}). Dashed lines are polynomial fits to the data.  Statistical uncertainties are of the order of the symbol size.} \label{one}
\end{figure}
Most of the results presented and discussed here pertain to adsorption of a \ph2 fluid inside a cylindrical channel of diameter $d$=10 \AA; the physical behaviour observed inside channels of greater or smaller diameter will be dealt with at the end.
We begin by discussing the energetics. Fig. \ref{one} shows computed values of the energy $e$ per \ph2 molecule as a function of the linear particle density $\nu=N/L$, for three of the alkali substrates considered here.
The results shown here are for a temperature $T$=1 K;  energy estimates for this system have been consistently found not to change significantly below $T=4$ K, at all physical conditions explored in this work.\\ \indent
The curves all display a minimum at the equilibrium density, corresponding to energy values all well below the bulk chemical potential of \ph2, namely $-90$ K. It is noteworthy that a stable adsorbate also exists inside a Cs channel (the energy minimum in that case is $-95$ K), which is in contrast to the case of an infinite flat Cs substrate, which is actually not wetted by \ph2.\cite{melting} Thus, the cylindrical geometry confers to the substrate greater adorption strength.
\begin{figure}
\centering
\setlength{\belowcaptionskip}{0pt}
\includegraphics[height=0.65\columnwidth]{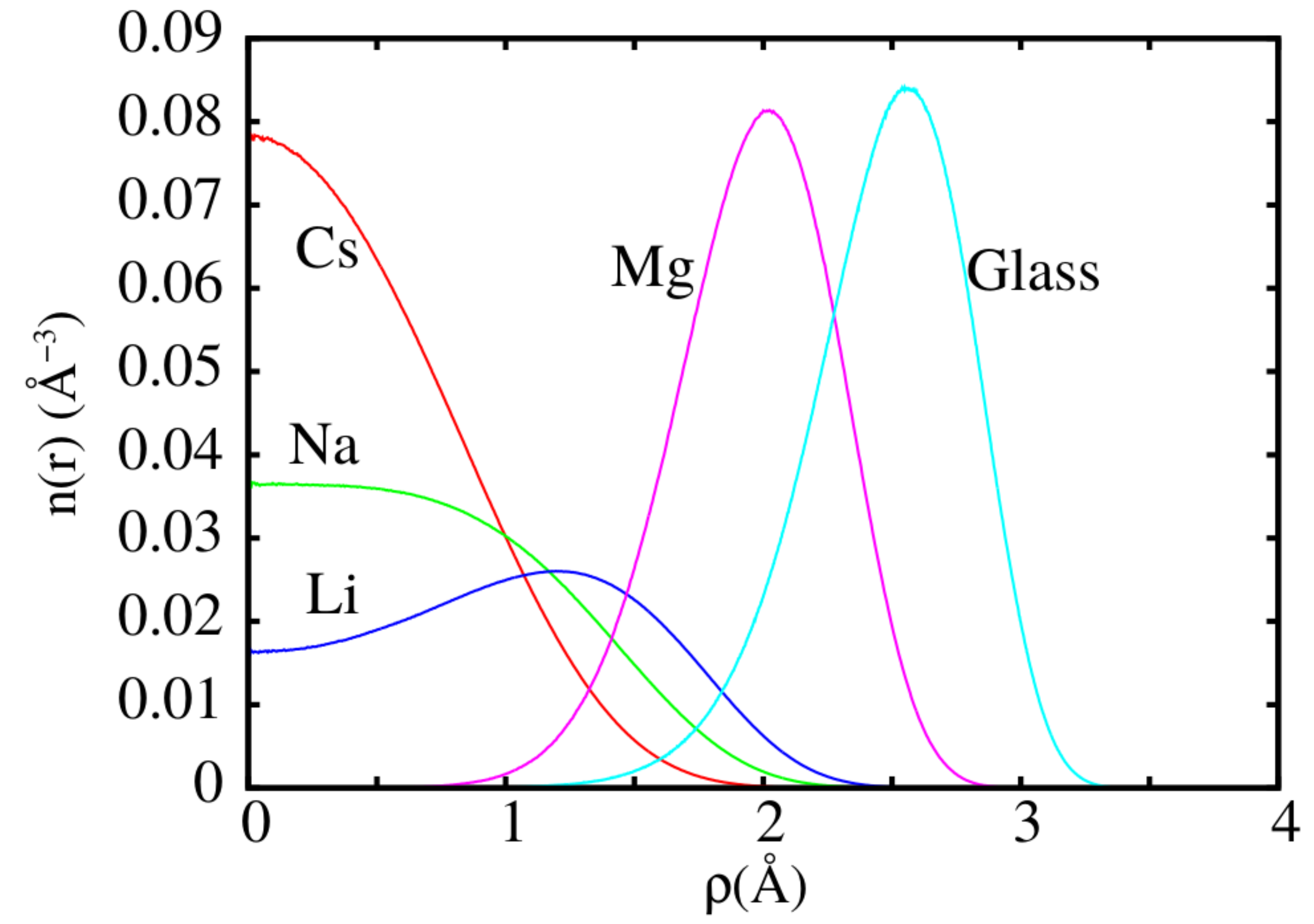}
\caption{{\it Color online}. Three-dimensional density of \ph2 inside a cylindrical channel of diameter $d$=10 \AA, plotted as a function of the distance from the axis. Curves pertain to different substrates, at the corresponding equilibrium densities. Statistical errors are not visible on the scale of the figure.} \label{two}
\end{figure}
It is seen that the equilibrium density increases with the strength of substrate adsorption, from a value close to 0.23 \AA$^{-1}$ for a K substrate (on a Cs it is essentially the same), which is $\sim 5$\% higher than the corresponding value for the purely 1D system,\cite{myprof4} to as high as 0.255 \AA$^{-1}$ inside a Li channel. These are three substrates for which the adsorbate displays a quasi-1D character, as shown by the \ph2 density profiles, computed with respect to the axis of the channel (Fig. \ref{two}). 
\\ \indent
On the two least attractive substrates among the five shown (Cs and Na), molecules line up along the axis of the cylinder, the greater pull that they experience in the case of a Na substrate resulting in a greater spread of the molecular wave function in the transverse direction. At the opposite end, on the stronger substrates such as Mg and Glass, the density of \ph2 is negligible on the axis, as the equilibrium phase consists of a single cylindrical shell coating the wall, with molecules sitting at a closer distance from it in the case of Glass.
\\ \indent
The physics of the system on the substrate labeled as Li, on the other hand, in a way interpolates between strong and weak adsorption.\cite{toigo} The density profile displays a maximum off the axis, but the density on the axis itself remains finite. This suggests that molecules arrange on a helix, winding around the axis; this structure remains largely 1D in character.
\begin{figure}[t]
\centering
\setlength{\belowcaptionskip}{0pt}
\includegraphics[height=2.4in]{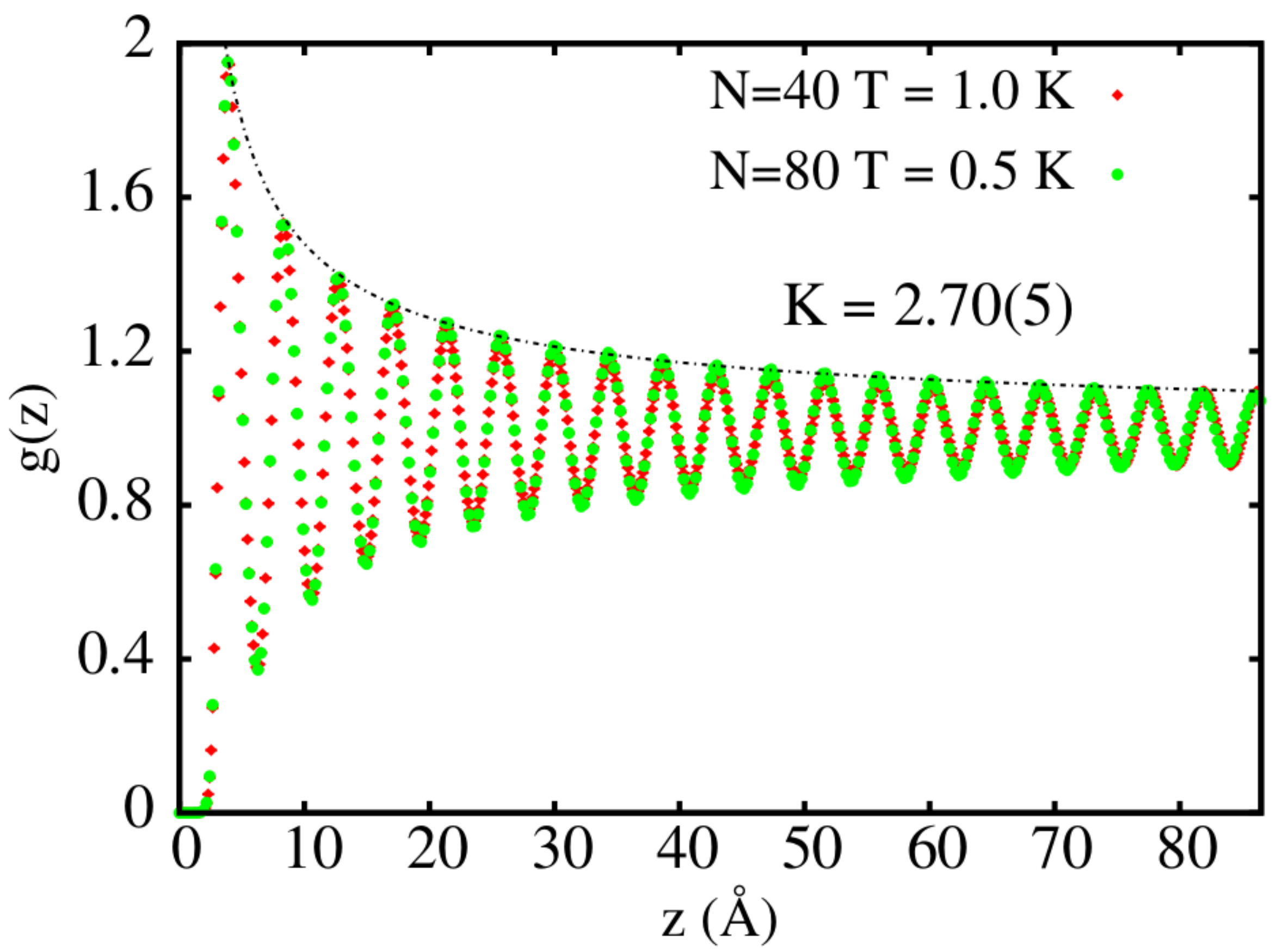}
\setlength{\belowcaptionskip}{0pt}
\includegraphics[height=2.4in]{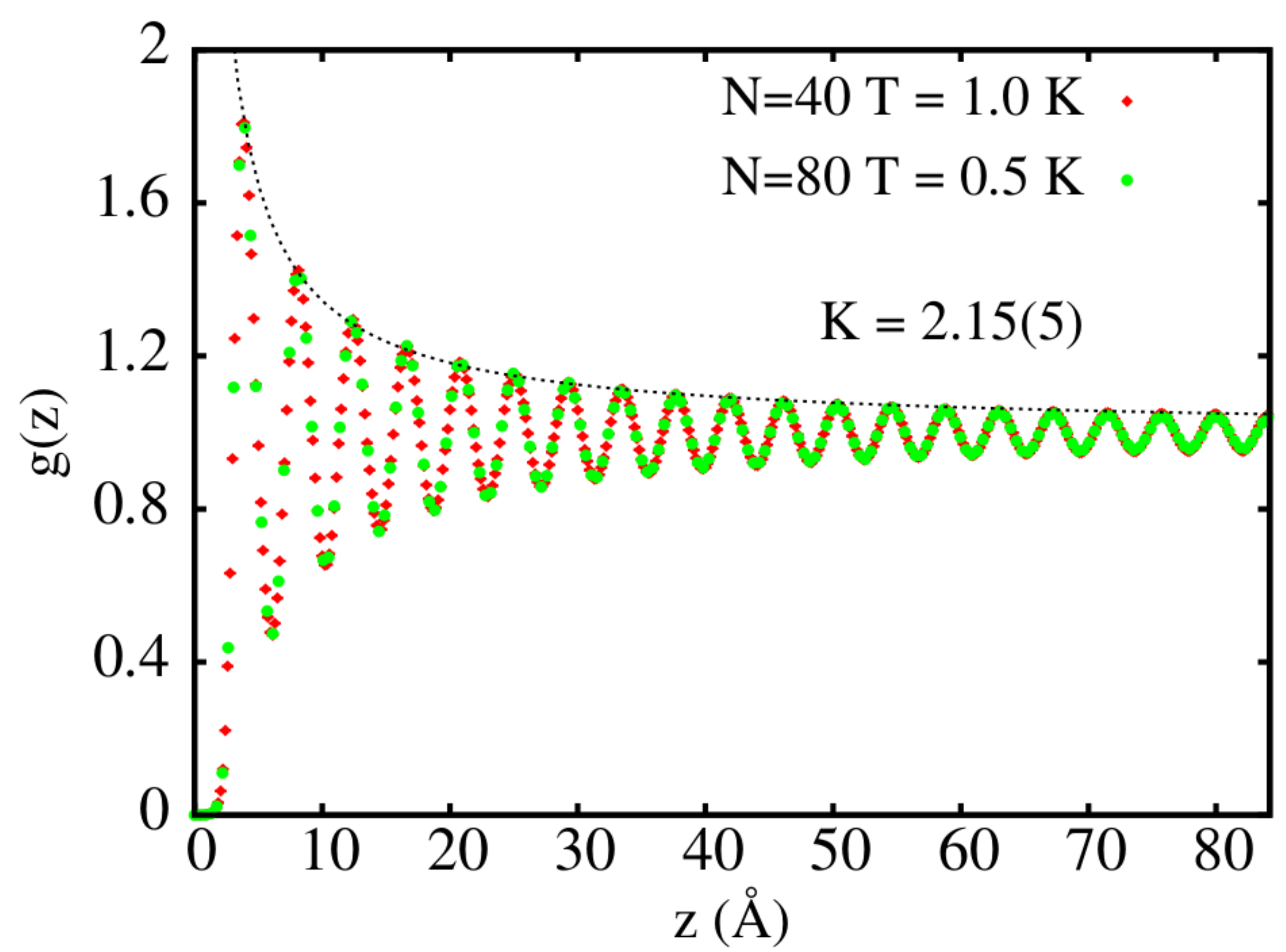}
\caption{{\it Color online}. Pair correlation functions
for \ph2 confined in a cylindrical channel of diameter $d$=10 \AA. Top panel shows results for 
a Cs substrate, bottom panel for a Na one. In both cases results are shown for the two temperatures $T$=1 K, for a system of $N=40$ molecules,  and $T$=0.5 K, for a system comprising $N=80$ molecules. Statistical errors are smaller of the sizes of the symbols.  Dashed lines represent fits to the maxima of $g(z)$ as explained in the text.
} 
\label{three}
\end{figure}
\\ \indent
An interesting question arises, namely,  what is the physical effect  of the significant molecular excursions away from the axis on the channel, on the physical character of a quasi-1D adsorbate. As mentioned in the Introduction, one may expect such excursions mainly to soften the intermolecular interaction, responsible for the strong propensity of the system to crystallize, possibly with the result of imparting to the system quasi-superfluid behaviour.
\\ \indent
In order to address this issue, we study the reduced pair correlation function $g(z)$, where $z$ is the component of the distance between two particles along the axis of the channel. For systems
that approach the 1D limit (and in our study that means equilibrium \ph2 layers adsorbed inside channels whose walls have the adsorption properties of the alkali metals in Table \ref{params}), the
$g(z)$ is expected {\it a}) to depend only  on the product $LT$ in the limit $L\to \infty, T\to 0$, and 
{\it b}) to conform to the behaviour predicted by Eq. \ref{maineq}, allowing one to infer\cite{myprof4,HeLutt} the value of the Luttinger parameter $K$ for the particular system of interest.
\\ \indent
Fig. \ref{three} shows pair correlation functions computed for \ph2 inside a Cs (top) and a Na (bottom) channel, in both cases for two different temperatures, namely $T=0.5$ and 1 K, and two different system sizes, comprising $N=40$ and 80 molecules. In both cases, calculations are carried out at the equilibrium density, which, as mentioned above, is slightly above that in purely 1D. As one can see, collapse of the data is clearly observed. The value of the parameter $K$ can be obtained by fitting the computed $g(z)$ to the expression (\ref {maineq}) or, somewhat more simply, its maxima to the expression $f(z)=1+A/z^{2/K}$.
\\ \indent
For the weakest substrate, for which the adsorbate is closest to the 1D limit, i.e. for which molecular excursions in the transverse direction are most limited, the determined value of $K$ is 2.70(5), appreciably lower than that for a purely 1D system (3.5). As the adsorption strength of the wall of the channel increases, our estimate of $K$ gradually decreases, the lowest value (very close to 2) attained for a Li substrate. As shown above (fig. \ref{two}), if the wall of the channel  is taken to be slightly more attractive than Li (i.e., Mg), then the equilibrium phase is a concentric cylindrical 
shell, with essentially no molecules in the central part of the channel. The effective 2D coverage of such a layer can be inferred from the data shown in Fig. \ref{two}, and is $\sim 0.067$ \AA$^{-2}$, i.e., the same with the equilibrium  coverage\cite{myprof3} of \ph2 in 2D at $T$=0.  Actually, the physics of such an adlayer is quite close to that of 2D \ph2; that is, the system displays solidlike behaviour, with molecules localized in space, quantum-mechanical exchanges are virtually absent and, consequently, no trace of superfluidity can be observed. On increasing the density, a second, quasi-1D inner layer eventually forms; we studied this system for the case of a glass channel. For simplicity, we utilized in these calculations an effective harmonic potential, adjusted to reproduce, in the vicinity of the axis of the cylinder, the combined effect of the interaction of the molecules with the wall of the channel, as well as with the molecules in the shell coating it. The resulting, fairly tight confining effect for the molecules in the central region, combined with the relatively high linear density of thermodynamically stable inner layers, imparts to the system in the inner part a markedly solidlike behaviour. Indeed, characteristic values of the parameter $K$ for quasi-1D inner \ph2 layers surrounded by a \ph2 cylindrical shell are generally $\ge 3.5$.
\\
Summarizing, the largest reduction of the Luttinger parameter $K$ that has been observed for \ph2 in cylindrical confinement, with respect to its value in 1D, while substantial (from 3.5 to 2), still leaves the system in the ``insulating" sector of the LLT.
\\ \indent
We now comment on the results obtained in narrower and wider channels. Inside a channel of diameter $d$=5 \AA, no adsorption occurs except for the most attractive of the substrates considered here, namely glass. The quasi-1D inner layer closely approaches the physics of 1D \ph2, with essentially the same value of the linear equilibrium density and Luttinger parameter $K$. Inside a channel of wider diameter ($d$=20 \AA), on the other hand, adsorption occurs for all substrates except Cs, and the equilibrium phase is again a single, solidlike cylindrical shell, concentric with the wall and with an equilibrium density close to that of 2D \ph2.
\section{CONCLUSIONS}\label{conc}
Extensive simulation studies have been performed for a realistic model of \ph2 adsorbed in the interior of a cylindrical channel of diameter ranging from 5 to 20 \AA. The results yield evidence that, although confinement can somewhat reduce the strong tendency of the system to crystallize, as observed in spherical cavities (Ref. \onlinecite{enhanced}), nevertheless the effect is quantitatively  more limited in a cylindrical geometry. 
\\ \indent
Specifically, if the diameter of the channel is as large as merely 2 nm, then the physics observed is qualitatively very similar to that which takes place when \ph2 is adsorbed on a flat substrate. On the other hand, inside narrow channels of diameter less than 1 nm, \ph2 will form quasi-1D adsorbate (if  the substrate is sufficienty strong) that closely reproduce the physics of the system in purely 1D, i.e., the phase is crystalline in nature (in the Luttinger sense).
\\ \indent
A cylindrical channel of diameter close to one nm, with a substrate that is relatively weak (e.g., Li)  provides a confining environment in which the interplay of reduced dimensionality and quantum excursions off the axis can lead to different physics, specifically to the stabilization of quasi-1D phases with a much reduced tendency to crystallize. This is qualitatively consistent with the recently reported\cite{enhanced} enhancement of the superfluid response of \ph2 clusters   trapped inside a spherical cavity; however, the effect is quantitatively far less significant in the quasi-1D geometry considered in this work, as the predicted reduction of the Luttinger parameter $K$ from its 1D value of 3.5 all the way to approximately 2 does not entail a fundamental change of the physical character of the system, which remains an insulator.\\ \indent
Consequently, any scheme aimed at stabilizing a bulk superfluid phase of \ph2 in a network of interconnected cavities seems to face the hurdle that no superflow may be sustained inside narrow cylindrical channels connecting two adjacent cavities.
\\ \indent
While this manuscript was undergoing review, we became aware of similar recent work,\cite{mah} claiming that the value of the Luttinger parameter
$K$ can be lowered considerably with respect to what found in this work, in fact to the point where \ph2 could turn superfluid (in the Luttinger sense),
inside Carbon nanotubes. The (cylindrical) geometry and diameters are similar to ours, but the model considered in Ref. \onlinecite{mah} includes substrate corrugation. These predictions, seemingly at variance with the results presented here,
were obtained with a different computational methodology with respect to that utilized here, specifically a ground state one. Clearly, further studies will be needed to resolve this discrepancy. 

\section*{Acknowledgments}
This work was supported by Natural Science and Engineering Research Council of Canada. Computing support of Westgrid is gratefully acknowledged.


\begin{thebibliography}{100}
\bibitem{expla}
The attractiveness of a wall potential depends both on the well depth $D$ and on the distance of closest approach $a$. In the following, when
speaking in general terms of ``attractive strength" of a substrate, or anything equivalent, we shall always refer to a particular combination of values of $D$ and $a$, not to the well depth $D$ alone.
\end{thebibliography}
\end{document}